\documentclass[12pt]{article}
\usepackage{graphicx}
\usepackage{longtable}
\begin{document}
\title{The light curves of type II-P SN 2017eaw: first 200 days}
\author{
        D.Yu.\,Tsvetkov$^1$,
        S.Yu.\,Shugarov$^{1,2}$,
       I.M.\,Volkov$^1$,\\
        N.N.\,Pavlyuk$^1$,
       O.V.\,Vozyakova$^1$,
       N.I.\,Shatsky$^1$,\\
      A.A.\,Nikiforova$^{3,4}$,
      I.S.\,Troitsky$^3$,
      Yu.V.\,Troitskaya$^3$,\\
      P.V.\,Baklanov$^{5,6,7}$
 }
\maketitle

   {\it
          $^1$Sternberg Astronomical Institute, M.V.\,Lomonosov Moscow State
          University,
          Universitetskii pr. 13, 119992 Moscow, Russia\\
          $^2$Astronomical Institute of the Slovak Academy of Sciences,
             059\,60 Tatransk\'{a} Lomnica, The Slovak Republic\\ 
          $^3$Astronomical Institute, St. Petersburg State University, 198504
          St. Petersburg, Russia\\
          $^4$Pulkovo Observatory, 196140 St.Petersburg, Russia\\
          $^5$Institute for Theoretical and Experimental Physics (ITEP), 117218
          Moscow, Russia\\
          $^6$ Novosibirsk State University, Novosibirsk 630090, Russia\\
          $^7$ National Research Nuclear University (MEPhI), Moscow 115409,
Russia}

\begin{abstract}
We present {\it UBVRI} photometry of the supernova 2017eaw in NGC\,6946,
obtained in the period from May 14 until December 7, 2017.
We derive dates and magnitudes of maximum light in the {\it UBVRI} bands and
the parameters of the light curves.
We discuss colour evolution, extinction and maximum luminosity of SN 2017eaw.
Preliminary modeling is carried out, and the results are in satisfactory
agreement with the  
light curves in the {\it UBVRI} bands.
\end{abstract}

\section*{Introduction}

Supernova (SN) 2017eaw was discovered 
by Patrick Wiggins on UT 2017 May 14.238 at magnitude 12.8
(Dong, Stanek, 2017). 
The SN was located at $\alpha=20^{\rm h}34^{\rm m}44^{\rm s}.238,
\delta=+60^{\circ}11'36''.00$, $61''.0$ west and $143''.0$ north of the 
center of NGC 6946 (Sarneczky et al., 2017)

The new object was classified as a young type II SN by Cheng et al. (2017)
and Tomasella et al. (2017).
The detection of the probable progenitor was reported by Khan (2017) and
Van Dyk et al. (2017).    

We present in this paper the results of extensive monitoring of SN\,2017eaw
in the optical bands undertaken at six observatories, and preliminary 
modeling of the light curves using the multi-group radiation-hydrodynamics
numerical code STELLA.

\section*{Observations}

Photometric observations  of SN\,2017eaw commenced on 2017 May 14,
immediately after discovery.
CCD frames in the 
{\it UBVRI} passbands were obtained at six sites, with 10 telescopes.
The telescopes used for monitoring are:
the 2.5-m telescope of the Caucasus Mountain Observatory of Sternberg
Astronomical Institute (SAI)(K250)(Potanin et al. 2017); 
the 1-m and 0.6-m telescopes of
Simeiz Observatory (S100, S60); the 0.7-m reflector of Crimean Astrophysical
Observatory (C70); the 0.6-m reflector of Crimean Observatory of SAI (C60);
the 0.7-m and 0.2-m telescopes of SAI in Moscow (M70, M20); the 0.6-m and 0.18-m 
telescopes of the Star\'a Lesn\'a Observatory
of the Astronomical Institute of the Slovak Academy of Sciences (T60, T18);
the 0.4-m telescope of the Astronomical Institute of
St.Petersburg State University (P40). 
 
All telescopes were equipped with CCD cameras and sets of Johnson-Cousins
{\it UBVRI} filters. 

\begin{figure}
\centerline{\includegraphics[width=12cm]{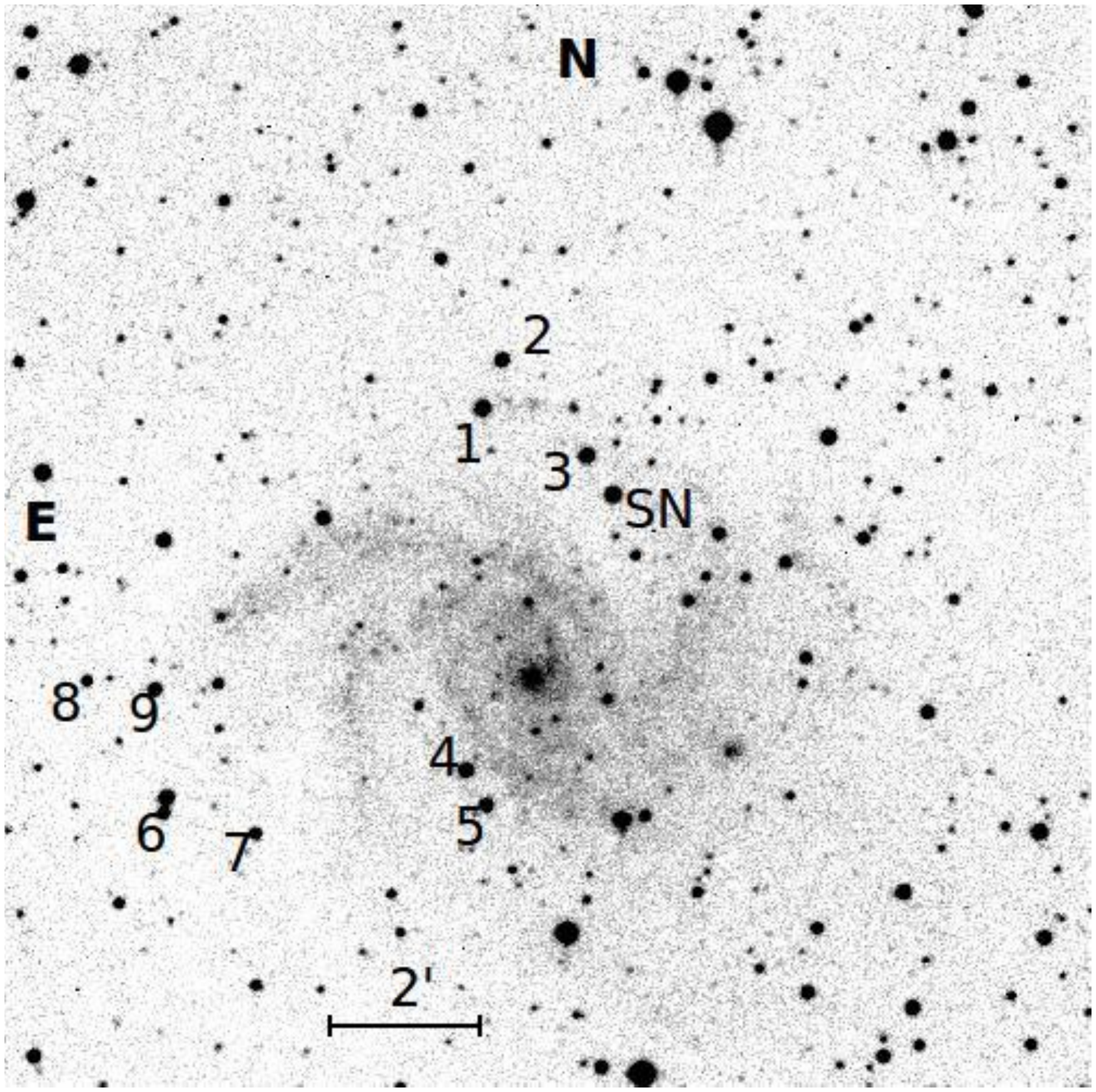}}
\caption{The image of SN 2017eaw and local standard stars, obtained
at the T60 telescope in the $V$-band}
\label{f1}  
\end{figure}

The standard image reductions and photometry were made using the
IRAF\footnotemark .
\footnotetext{IRAF is distributed by the National Optical 
Astronomy Observatory,
which is operated by AURA under cooperative agreement with the
National Science Foundation.}
The magnitudes of the SN
were derived by aperture photometry or PSF-fitting relatively to
local standard stars.
The CCD image of SN 2017eaw and local standard stars is presented
in Fig.\,\ref{f1}.
The magnitudes of stars 4,5 were reported by Botticella et al. (2009),
stars 6--9 were calibrated by Misra et al. (2007).
Some of the instruments employed for photometry have small field of 
view of about $6'$ (M70, S60), and only stars 1--3 could be measured on 
the frames. We calibrated the stars 1--3 on images with larger field of view
relative to the stars 4--9.  

The surface brightness of the host galaxy at the location of the SN
is low, and subtraction of galaxy background is not necessary.

The photometry was transformed to the standard Johnson-Cousins
system by means of instrument colour-terms, determined from
observations of standard star clusters. 

Our photometry of the SN is presented
in Table\,1.

\begin{center}
\small
\label{t1}
\begin{longtable}[t]{cccccccccccl}
\caption{{\it UBVRI} magnitudes of SN\,2017eaw}\\ 
\hline\hline
JD$-$ & $U$ & $\sigma_U$ & $B$ & $\sigma_B$ & $V$ & $\sigma_V$ &
$R$ & $\sigma_R$ & $I$ & $\sigma_I$& Tel. \\
2450000 & & & & & & & & & & &\\
\hline
\endfirsthead
\caption{Continued.}\\
\hline
JD$-$ & $U$ & $\sigma_U$ & $B$ & $\sigma_B$ & $V$ & $\sigma_V$ &
$R$ & $\sigma_R$ & $I$ & $\sigma_I$& Tel. \\
2450000 & & & & & & & & & & &\\
\hline  
\endhead
\hline  
\endfoot
\hline\hline
\endlastfoot
\small
7888.47& 12.62 &0.04& 13.32& 0.02& 13.13& 0.01& 12.92& 0.02& 12.68& 0.02&T60\\
7890.41&       &    & 13.30& 0.04& 12.93& 0.02& 12.64& 0.02& 12.37& 0.03&P40 \\
7892.41& 12.53 &0.06& 13.25& 0.03& 12.89& 0.02& 12.54& 0.02& 12.19& 0.03&M70\\ 
7892.52&       &    & 13.20& 0.02& 12.86& 0.01& 12.54& 0.02& 12.18& 0.02&S100\\
7898.52& 12.50 &0.09& 13.33& 0.03& 12.97& 0.02& 12.56& 0.02& 12.22& 0.03&C60\\ 
7901.45& 12.94 &0.01&      &     & 12.99& 0.01& 12.58& 0.01& 12.26&0.01&K250 \\
7901.48& 12.82 &0.10& 13.37& 0.03& 13.00& 0.01& 12.57& 0.03& 12.23& 0.03&C60\\ 
7904.35& 13.10 &0.06& 13.48& 0.03& 13.05& 0.01& 12.55& 0.02& 12.16& 0.03&M70\\ 
7904.46& 13.20 &0.01&      &     & 13.01& 0.01& 12.59& 0.01& 12.26&0.01&K250\\  
7905.49& 13.22 &0.06& 13.52& 0.02& 13.02& 0.02& 12.58& 0.02& 12.24& 0.03&
C60\\ 
7912.38&       &    & 13.82& 0.02& 13.06& 0.01& 12.56& 0.02& 12.15& 0.02&
M70 \\
7912.39&       &    & 13.79& 0.03& 13.00& 0.02& 12.53& 0.02& 12.16& 0.03&
S100 \\
7914.45&       &    & 13.93& 0.04& 13.10& 0.02& 12.59& 0.03& 12.21& 0.04&
T18 \\ 
7916.39& 14.32 &0.04& 14.01& 0.02& 13.11& 0.01& 12.64& 0.02& 12.26& 0.02&
T60 \\ 
7916.47&       &    & 14.00& 0.04& 13.07& 0.03& 12.64& 0.03& 12.24& 0.03&
C70 \\ 
7921.53&       &    & 14.18& 0.03& 13.18& 0.01& 12.67& 0.02& 12.24& 0.02&
S100 \\
7923.45& 14.80 &0.04& 14.22& 0.02& 13.19& 0.02& 12.69& 0.02& 12.27& 0.02&
T60 \\ 
7926.31&       &    & 14.31& 0.02& 13.21& 0.01& 12.69& 0.02& 12.26& 0.02&
S100 \\
7926.45&       &    & 14.26& 0.05& 13.20& 0.02& 12.70& 0.03& 12.25& 0.04&
C70 \\ 
7927.54&       &    & 14.34& 0.02& 13.22& 0.02& 12.70& 0.02& 12.27& 0.02&
S100\\ 
7928.41&       &    & 14.35& 0.03& 13.22& 0.01& 12.69& 0.02& 12.19& 0.02&
M70 \\ 
7929.34& 15.04 &0.05& 14.35& 0.03& 13.22& 0.01& 12.71& 0.02& 12.26& 0.03&
T60 \\ 
7930.33&       &    & 14.40& 0.02& 13.23& 0.02& 12.72& 0.02& 12.24& 0.03&
S100 \\
7931.32&       &    & 14.41& 0.02& 13.24& 0.01& 12.71& 0.02& 12.25& 0.02&
S100 \\
7932.30&       &    & 14.39& 0.04& 13.24& 0.01& 12.71& 0.02& 12.24& 0.02&
S100 \\
7933.31&       &    & 14.44& 0.02& 13.23& 0.02&      &     & 12.24& 0.03&
S100 \\
7934.30&       &    & 14.46& 0.03& 13.22& 0.02&      &     & 12.24& 0.04&
S100 \\
7934.43& 15.25 &0.05& 14.42& 0.03& 13.23& 0.02& 12.71& 0.02& 12.24& 0.02&
T60  \\
7940.32&       &    & 14.52& 0.02& 13.27& 0.02& 12.71& 0.02& 12.22& 0.02&
S100 \\
7942.40&       &    & 14.56& 0.02& 13.28& 0.01& 12.72& 0.02& 12.23& 0.02&
S100 \\
7942.42&       &    & 14.47& 0.04& 13.26& 0.03& 12.69& 0.04& 12.20& 0.05&
C70 \\ 
7944.55&       &    & 14.59& 0.03& 13.28& 0.02& 12.71& 0.02& 12.21& 0.02&
S100 \\
7956.46&       &    & 14.68& 0.03& 13.30& 0.03& 12.72& 0.03& 12.23& 0.03&
C70 \\ 
7960.50& 16.14 &0.05& 14.75& 0.02& 13.33& 0.02& 12.71& 0.02& 12.21& 0.02&
T60  \\
7966.33&       &    &      &     & 13.38& 0.02& 12.73& 0.02& 12.21& 0.03&
M20  \\
7967.56& 16.31 &0.04& 14.82& 0.02& 13.37& 0.02& 12.73& 0.02& 12.23& 0.02&
T60  \\
7972.28&       &    & 14.92& 0.02& 13.41& 0.02& 12.79& 0.03& 12.22& 0.06&
S100 \\
7975.57&       &    & 14.90& 0.05& 13.43& 0.02& 12.79& 0.03&      &     &
S60  \\
7978.50&       &    & 15.05& 0.02& 13.51& 0.01& 12.80& 0.02& 12.27& 0.03&
S100 \\
7980.47&       &    & 15.11& 0.03& 13.55& 0.01& 12.83& 0.02& 12.30& 0.02&
S100 \\
7981.36&       &    & 15.10& 0.04& 13.58& 0.02& 12.87& 0.02& 12.32& 0.03&
M20  \\
7982.53& 16.91 &0.05& 15.16& 0.02& 13.59& 0.01& 12.88& 0.02& 12.34& 0.02&
T60  \\
7985.54&       &    & 15.18& 0.02& 13.63& 0.02& 12.89& 0.02& 12.36& 0.03&
S100 \\
7986.29&       &    & 15.23& 0.08& 13.67& 0.02& 12.92& 0.04& 12.38& 0.03&
M20  \\
7990.37&       &    & 15.42& 0.03& 13.80& 0.02& 13.02& 0.03&      &     &
S60  \\
7990.48& 17.24 &0.07& 15.45& 0.04& 13.81& 0.01& 13.05& 0.02& 12.49& 0.02&
T60  \\
7991.38&       &    & 15.44& 0.02& 13.86& 0.01& 13.09& 0.02& 12.52& 0.03&
C60  \\
7992.49&       &    & 15.48& 0.02& 13.88& 0.01& 13.11& 0.02& 12.53& 0.02&
C60  \\
7992.58&       &    & 15.54& 0.04& 13.88& 0.01& 13.11& 0.01& 12.51& 0.03&
S60  \\
7993.53&       &    & 15.53& 0.02& 13.94& 0.02& 13.13& 0.02& 12.55& 0.02&
C60  \\
7994.46&       &    & 15.60& 0.03& 13.99& 0.01& 13.18& 0.02& 12.60& 0.02&
C60  \\
7995.37&       &    &      &     & 13.99& 0.02& 13.22& 0.03& 12.61& 0.03&
P40 \\ 
7995.45& 17.69 &0.06& 15.76& 0.03& 14.02& 0.01& 13.22& 0.03& 12.62& 0.03&
T60  \\
7997.30&       &    & 15.74& 0.04& 14.07& 0.02& 13.26& 0.03& 12.68& 0.03&
P40 \\ 
7997.33&       &    & 15.87& 0.05& 14.15& 0.03& 13.32& 0.03& 12.69& 0.04&
M20 \\ 
7998.36&       &    & 15.81& 0.03& 14.14& 0.02& 13.34& 0.02& 12.81& 0.03&
P40 \\ 
7999.44&       &    & 15.84& 0.05& 14.14& 0.05&      &     & 12.87& 0.06&
C70 \\ 
8001.55&       &    &      &     &      &     & 13.53& 0.02& 12.97& 0.03&
P40 \\ 
8004.29&       &    & 16.48& 0.02& 14.77& 0.01& 13.80& 0.02& 13.14& 0.02&
C60 \\ 
8005.37&       &    & 16.61& 0.02& 14.90& 0.01& 13.93& 0.02& 13.25& 0.02&
C60 \\ 
8006.32&       &    & 16.75& 0.02& 15.03& 0.01& 14.03& 0.02& 13.34& 0.02&
C60 \\ 
8007.26&       &    & 16.88& 0.03& 15.15& 0.01& 14.14& 0.02& 13.45& 0.02&
C60 \\ 
8008.30&       &    & 17.06& 0.03& 15.28& 0.01& 14.26& 0.02& 13.57& 0.03&
C60 \\ 
8008.38&       &    &      &     &      &     & 14.23& 0.02& 13.58& 0.04&
P40 \\ 
8009.25&       &    & 17.09& 0.02& 15.38& 0.01& 14.33& 0.02& 13.64& 0.02&
C60 \\ 
8009.32&       &    & 17.03& 0.13& 15.37& 0.05& 14.39& 0.03& 13.67& 0.04&
M20 \\ 
8009.47&       &    & 17.12& 0.08& 15.27& 0.05& 14.27& 0.04&      &     &
C70 \\ 
8011.44&       &    & 17.22& 0.04& 15.43& 0.08& 14.37& 0.03& 13.79& 0.04&
C70 \\ 
8013.43&       &    &      &     & 15.45& 0.05& 14.45& 0.02&      &     &
C70 \\ 
8014.31&       &    &      &     & 15.49& 0.02& 14.52& 0.02& 13.87& 0.03&
P40 \\ 
8019.19&       &    &      &     & 15.72& 0.04& 14.66& 0.03& 13.99& 0.06&
M20 \\ 
8019.45&       &    & 17.42& 0.07& 15.62& 0.04& 14.61& 0.02& 13.91& 0.03&
P40 \\ 
8020.44&       &    &      &     &      &     & 14.61& 0.02& 13.97& 0.03&
P40 \\ 
8021.28&       &    &      &     & 15.64& 0.05& 14.65& 0.03& 13.97& 0.03&
P40 \\ 
8022.30&       &    & 17.59& 0.03& 15.74& 0.01& 14.66& 0.02& 13.94& 0.02&
T60 \\ 
8023.21&       &    &      &     & 15.72& 0.06& 14.65& 0.03&      &     &
M70 \\ 
8025.38&       &    & 17.44& 0.06& 15.73& 0.05& 14.64& 0.05& 13.93& 0.03&
C70\\  
8027.35&       &    & 17.51& 0.10& 15.72& 0.06& 14.65& 0.05& 13.94& 0.04&
C70\\  
8028.43&       &    & 17.63& 0.03& 15.80& 0.02& 14.72& 0.03& 13.98& 0.03&
T60 \\ 
8029.40&       &    & 17.43& 0.09& 15.71& 0.06& 14.63& 0.03& 13.96& 0.03&
C70\\  
8040.36&       &    &      &     &      &     & 14.80& 0.02& 14.20& 0.03&
P40\\  
8044.15&       &    &      &     & 15.82& 0.09& 14.78& 0.04& 14.16& 0.04&
M70 \\ 
8049.40&       &    &      &     &      &     & 14.85& 0.02& 14.23& 0.03&
P40\\  
8052.30&       &    & 17.74& 0.05& 16.00& 0.02& 14.91& 0.02& 14.23& 0.03&
T60 \\ 
8059.47&       &    &      &     &      &     & 14.94& 0.02& 14.34& 0.03&
P40\\  
8061.15&       &    & 17.72& 0.05& 16.16& 0.03& 15.03& 0.04& 14.37& 0.04&
C60 \\ 
8063.14&       &    & 17.81& 0.03& 16.18& 0.01& 15.04& 0.02& 14.34& 0.02&
C60 \\ 
8074.22&       &    &      &     & 16.31& 0.02& 15.12& 0.03&      &     &
C60 \\ 
8075.27&       &    & 17.81& 0.04& 16.33& 0.02& 15.15& 0.02& 14.45& 0.02&
C60 \\ 
8076.20&       &    & 17.87& 0.03& 16.34& 0.02& 15.16& 0.02& 14.47& 0.02&
C60\\   
8095.22&       &    & 18.03& 0.04& 16.43& 0.02& 15.32& 0.02& 14.60& 0.02&
T60\\    
\end{longtable}
\end{center}

\section*{Light and colour curves}

\begin{figure}
\centerline{\includegraphics[width=12cm]{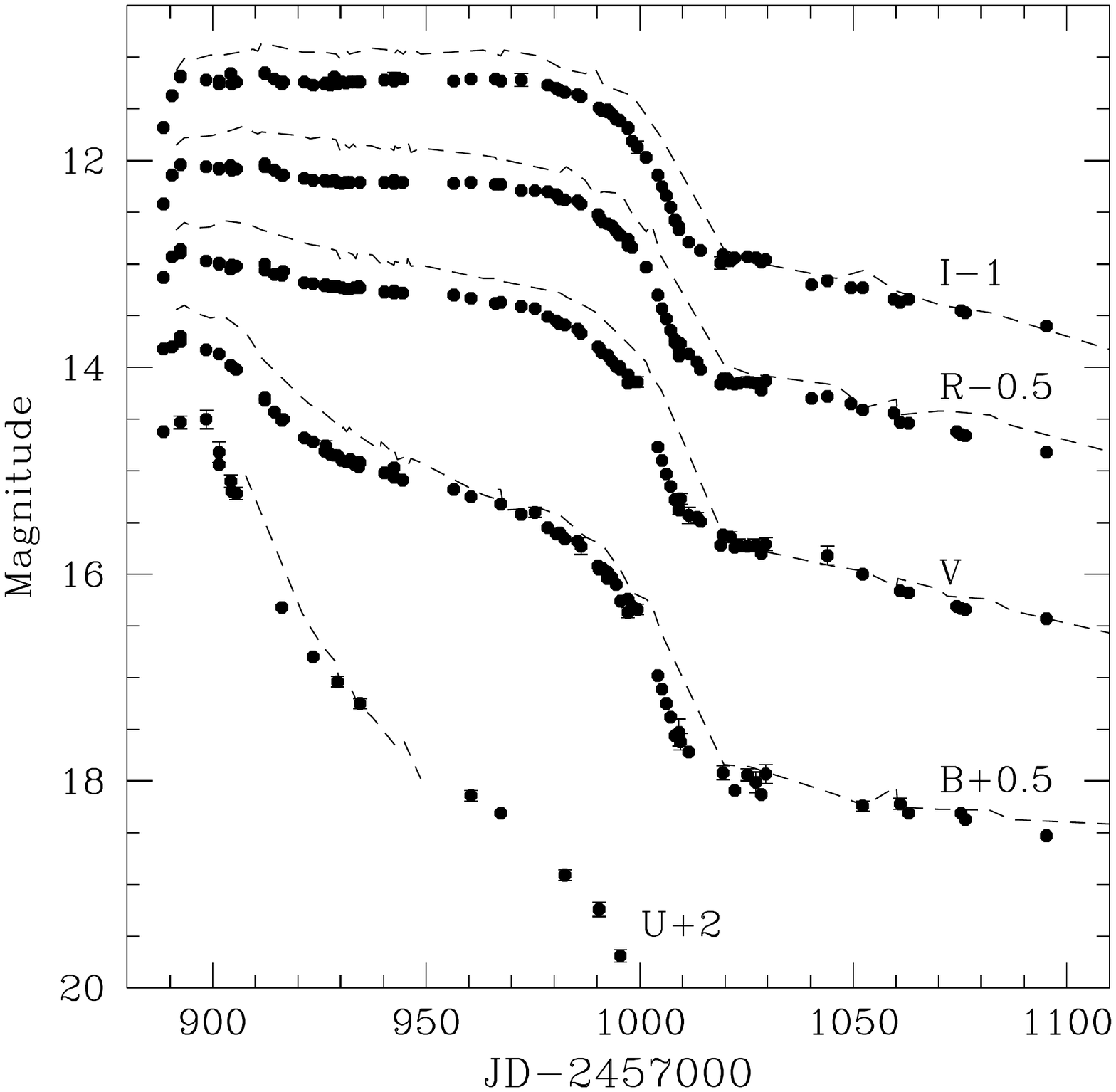}}
\caption{The light curves of SN\,2017eaw in the {\it UBVRI} bands.
The error bars are plotted only when they exceed the size of
a symbol. The dashed lines show the light curves of SN\,2004et.
The curves are shifted for clarity, the amount of shift is indicated on 
the plot}
\label{f2}
\end{figure}

The light curves of SN\,2017eaw are presented in Fig.\,\ref{f2}.
The results for all the telescopes are in a
fairly good agreement. The light curves are typical for type II-P 
SNe, with a plateau lasting about 100 days. Our first observations
were made before maximum light, and the date of maximum can be determined
as JD\,2457892.5. The magnitudes at maximum are $U=12.5, B=13.22, V=12.87,
R=12.54, I=12.18$.
The fast decline after the plateau stage started at about JD\,2457980, and 
the final linear tail started at JD\,2457810, with $B=17.3, V=15.5, R=14.5, 
I=13.8$ mag. The rates of decline on the tail in
the $B,V,R,I$ bands are, respectively, 0.0054, 0.0106, 0.0092,
0.0090 mag\,day$^{-1}$. 

The light curves of SN\,2004et (Maguire et al., 2010), 
which occured in the same galaxy and also
belonged to the type II-P, are plotted for comparison. We shifted the curves 
for SN\,2004et only in time, assuming JD\,2453270.5 as the date of explosion 
for SN\,2004et (Maguire et al., 2010), and JD\,2457884 as explosion date
for SN\,2017eaw (Tomasella et al., 2017). The shape of the light curves
is practically identical, but SN\,2004et is about 0.3 mag
britghter at the plateau stage in the $B,V,R,I$ bands. But on the tail the
difference between two SNe becomes negligible.

\begin{figure}
\centerline{\includegraphics[width=12cm]{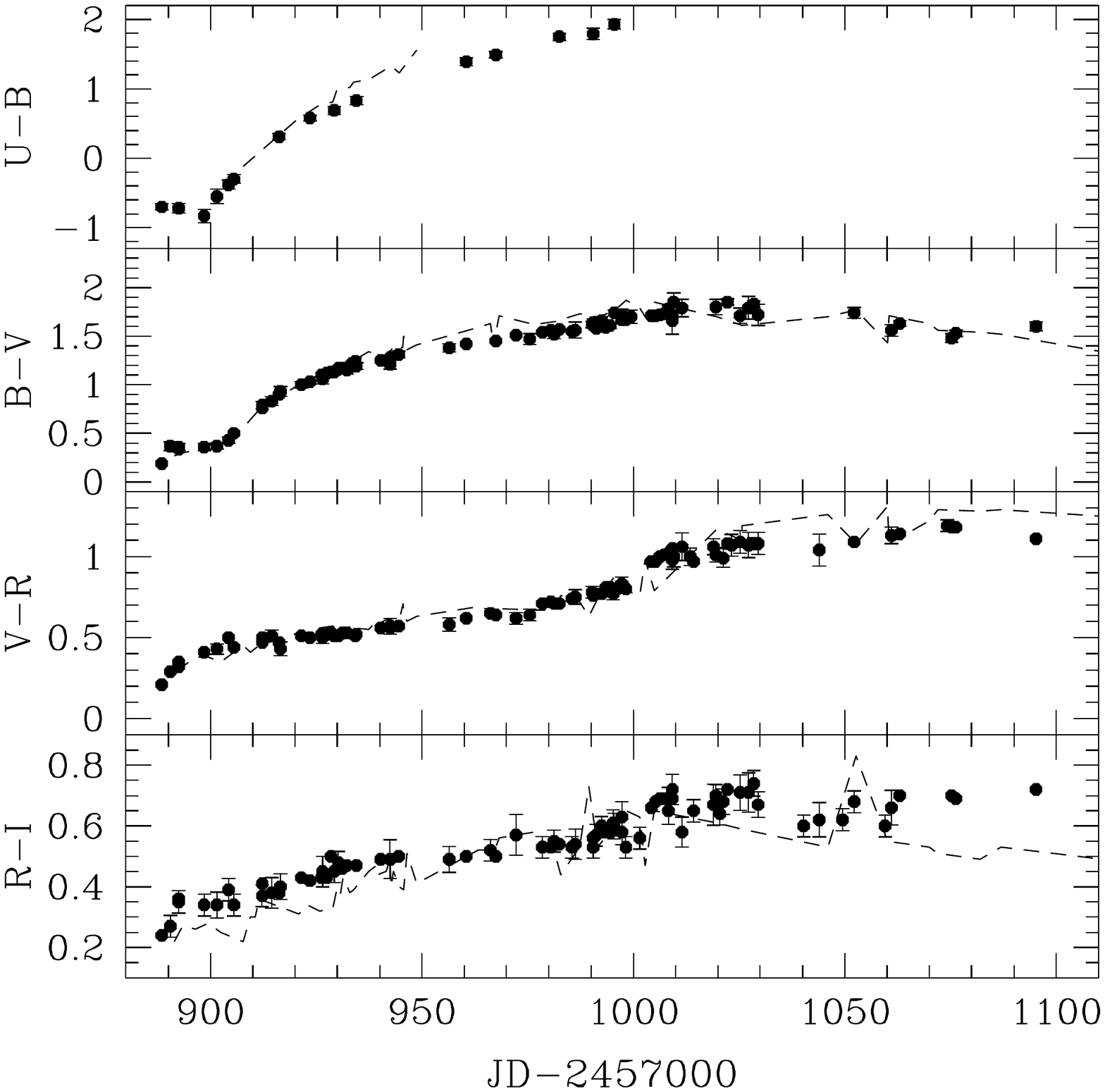}}
\caption{The colour curves of SNe 2017eaw, 
the dashed lines show 
the colour curves for SN 2004et}
\label{f3}
\end{figure}

The colour curves for SNe 2004et and 2017eaw are compared in Fig.\,\ref{f3}.
The similarity of the curves is evident, 
some difference can be noticed only for the $R-I$ colour.
We may suppose that the values
of interstellar extinction for these SNe are very close. 
The reddening for SN\,2004et was estimated as $E(B-V)=0.41$ mag by Maguire et
al. (2010) considering the strength of interstellar Na\,I lines. Using the 
same method, Tomasella et al. (2017) derived $E(B-V)=0.22$ mag for SN\,2017eaw.
The Galactic extinction in the direction of NGC\,6946 is $E(B-V)=0.30$
mag according to Schlafly, Finkbeiner (2011) (via NED\footnotemark).
\footnotetext{http://ned.ipac.caltech.edu}
We suppose, that the extinction for both SNe has nearly equal value 
and occurs only in 
the Galaxy. The difference of estimates based on the equivalent width 
of Na I lines may be
due to the large dispersion of results obtained with this method.      
This supposition is supported by the fact that both SNe exploded at 
quite large distances from the center of the galaxy: 8 kpc for SN\,2004et,
4.7 kpc for SN\,2017eaw, in the regions with low surface brightness.

Assuming the distance of 6.0 Mpc for NGC\,6946 (Efremov et al., 2011) and 
reddening $E(B-V)=0.30$ mag we derive absolute magnitudes of SN\,2017eaw 
at maximum light: $M_U=-17.9, M_B=-16.9, M_V=-17.0, M_R=-17.1, M_I=-17.2$
mag. 
The value of $M_B$ is only slightly brighter than the mean absolute
magnitude $M_B=-16.75$ mag for SNe II-P (Richardson et al., 2014).
SN\,2004et is about 0.3 mag brighter, but its luminosity is 
also quite close to the mean value for SNe II-P. 

\section*{Modeling}

The light curve shapes with a luminous plateau shows that SN\,2017eaw is
a normal type II-P SN and its presupernova star was a red supergiant (RSG).
SNe II-P show large variety in their light-curve shapes. 
The main features of the light curves are determined by the initial radius $R$, 
total mass of presupernova $M$, mass of $^{56}$Ni and the energy of 
explosion $E$
(Litvinova, Nadyozhin, 1985; Kasen, Woosley, 2009).
We computed the large grid of models on  parameter space ($R,M$,$^{56}$Ni,
$E$) to
evaluate the best fit model.

For the model calculation, we use the multi-group radiation-hydrodynamics
numerical code STELLA 
(Blinnikov et al., 1998; 2000; 2006) 

We constructed the presupernova RSG models in non-evolutionary hydrostatic
equilibrium described previously (Baklanov et al., 2005).
The shock propagation  causes strong mixing of envelope matter due to
Rayleigh-Taylor instability.
The amount and distribution of $^{56}$Ni are manually adjusted 
to make its radial distribution closer to the actual distribution.  
The density and the elements distribution for the presupernova model 
R600M23Ni005E20
are shown in Fig. 4.
 \begin{figure}
  \begin{center}
  \includegraphics[width=12cm]{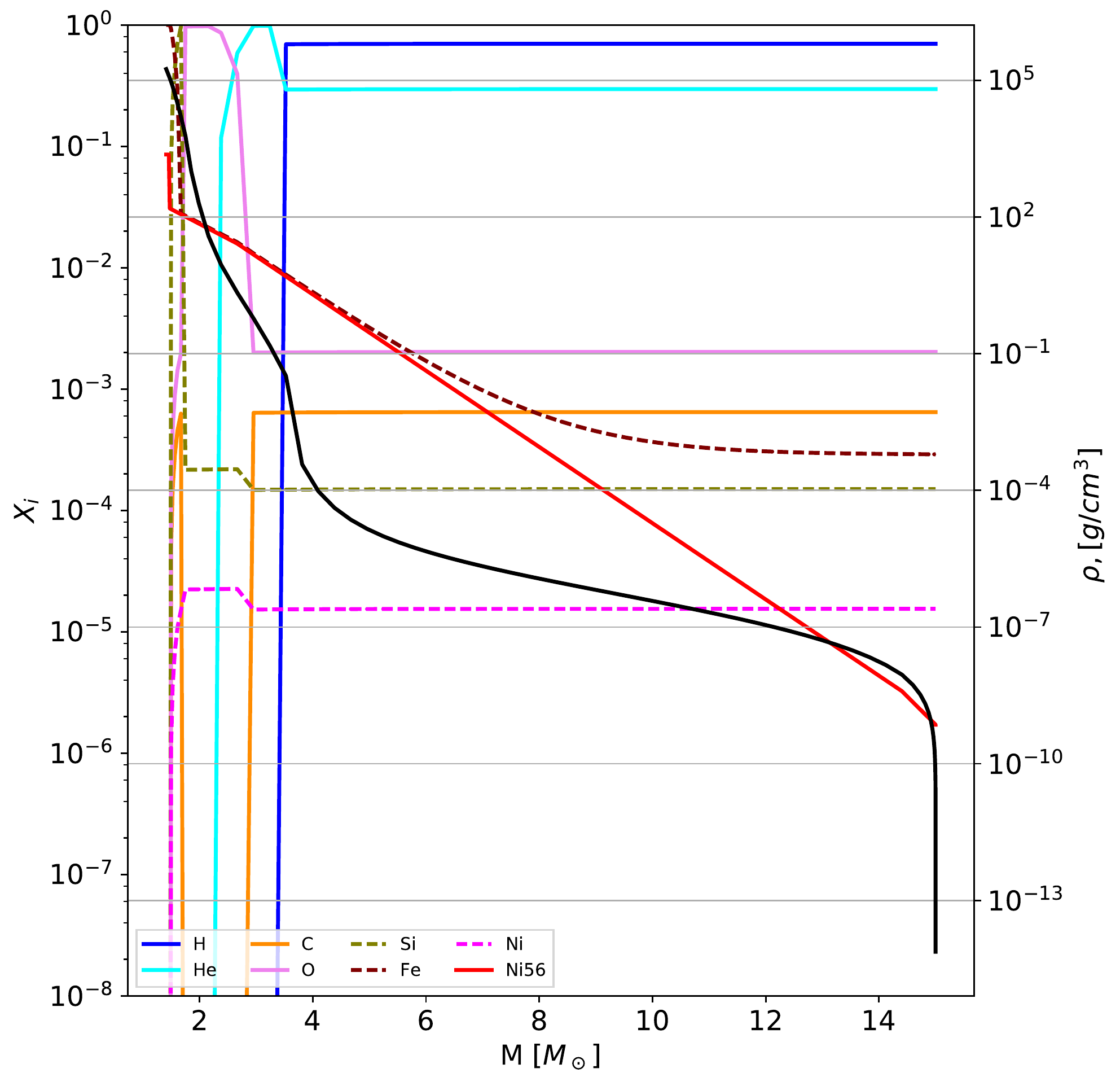}
  \end{center}
  \caption{The density and the elements distribution in the presupernova for
  the model R600M23Ni005E20}
  \label{f4}
 \end{figure}
The SN explosion was simulated by the release of $E_{exp} =
2\times10^{51}\,$erg = 2\,foe in the form of a ``thermal bomb'' 
in the innermost region of the ejecta. The parameters of the 
model R600M23Ni005E20 are $R=600R_{\odot}, M=23M_{\odot}, 
M_{\rm Ni}=0.05M_{\odot}$. 
 
We found that the light curves of SN\,2017eaw
are better reproduced with the supernova model R600M23Ni005E20.
(Fig. 5). 
The model R1100M15Ni005E9 has longer interval between explosion 
and maximum light, which does not
agree with the observations (Fig. 6).

 \begin{figure}
   \begin{center}
   \includegraphics[width=12cm]{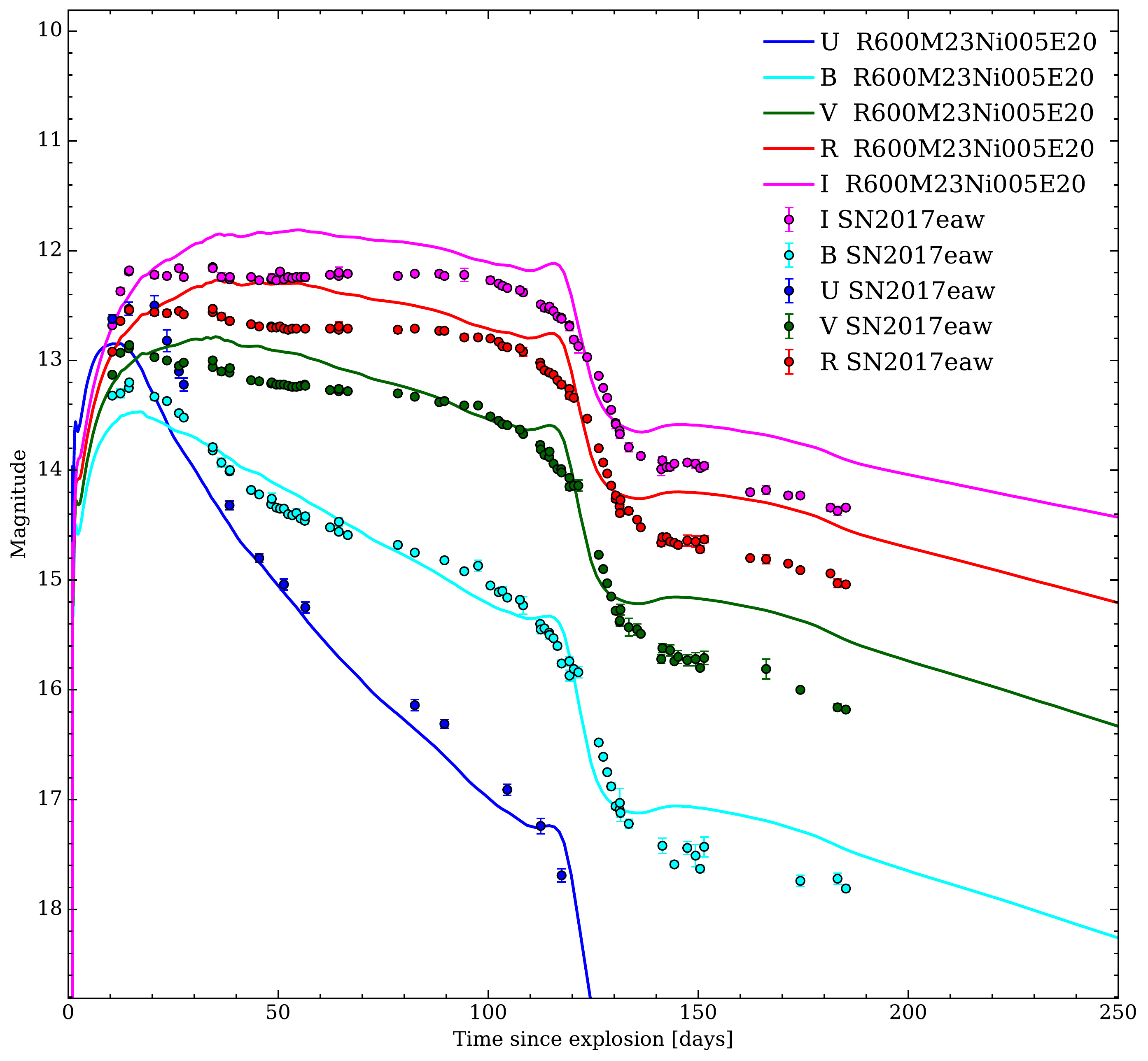}
   \end{center}
   \caption{{\it UBVRI} light curves of SN\,2017eaw for the model
R600M23Ni005E20. 
Time from the explosion is along the horizontal axis.}
   \label{f5}
  \end{figure}
  
This is a preliminary result, which requires further investigation when more
information is available, in particular on the rate of expansion of the
supernova envelope.
But it is remarkable that the our simulation results are in good agreement 
with the results of independent calculations of Utrobin and Chugai (2009) 
for SN\,2004et, which earlier exploded in the same galaxy NGC\,6946.
 
\begin{figure}
  \begin{center}
  \includegraphics[width=12cm]{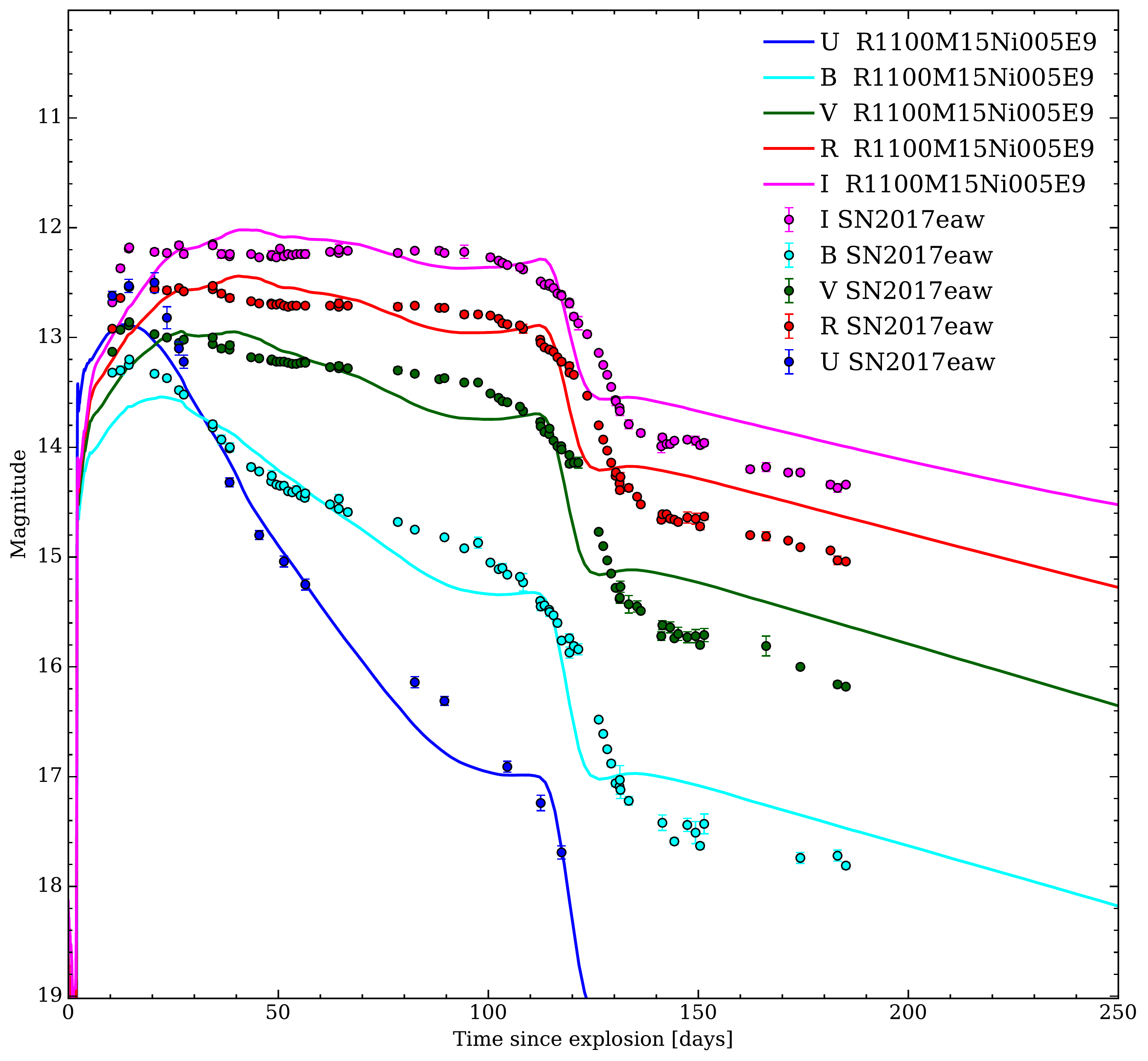}
  \end{center}
  \caption{{\it UBVRI} light curves of SN\,2017eaw for the model
 R1100M15Ni005E9}
  \label{f6}
 \end{figure}

\section*{Conclusions}

We present the light and colour curves of SN\,2017eaw. Our observations 
started immediately after discovery, and we detected the rising part
of the light curves. The photometric evolution was followed through the
plateau stage and to the linear tail. We determined the basic parameters
of the light curves and estimated the maximim absolute magnitudes. 
The shape of the light curves and maximum luminosity of SN\,2017eaw are
typical for the class II-P.
We compared the light curves of SN\,2017eaw with those for SN\,2004et which
exploded im the same galaxy and also belonged to the type II-P. Two 
objects have very similar photometric evolution and we supposed that 
interstellar reddening for them is nearly equal.

We present preliminary results of model calculation using the multi-group
radiation-hydrodynamics numerical code STELLA.  
 
We continue the observations of SN\,2017eaw, the results and more detailed
analysis of the data will be presented in a subsequent paper.

\section*{Acknowledgements}

The work of D.Tsvetkov and P.Baklanov was partly supported 
by the Russian Science
Foundation Grant No. 16-12-10519. 
The work of S.Shugarov was partially supported by Grants
VEGA 2/0008/17 and APVV-15-0458.

This work was performed with the equipment purchased from the funds of
the Program of
Development of Moscow University.

I.Volkov was partly supported by the Russian Science
Foundation Grant No. 14-12-00146.
 
 
The authors are grateful to I.V.Nikolenko for help with the observations
and to T.S.Grishina, S.S.Savchenko, E.G.Larionova, E.N.Kopatskaya,
G.A.Borman and A.A.Vasilyev, who carried out some of the observations.

\end{document}